\documentclass[%
 reprint,
superscriptaddress,
showpacs,preprintnumbers,
nofootinbib,
 amsmath,amssymb,
 bm,
 aps,
]{revtex4-1}

\usepackage{graphicx}
\usepackage{amssymb}
\usepackage{epstopdf}
\usepackage{graphicx}
\usepackage{dcolumn}
\usepackage{bm}
\usepackage{color}
\usepackage[caption=false]{subfig}
\usepackage{mhchem}

\newcommand{\be}{\begin{eqnarray}}

\newcommand{\ee}{\end{eqnarray}}

\usepackage[normalem]{ulem}

\usepackage[export]{adjustbox}

\providecommand{\sorthelp}[1]{}

\begin{document}

\preprint{APS/123-Q ED}

\title{The primordial information content of Rayleigh Anisotropies}

\newcommand{\IPMU}{Kavli Institute for the Physics and Mathematics of the Universe (WPI), The University of Tokyo Institutes for Advanced Study (UTIAS), The University of Tokyo, Kashiwa, Chiba 277-8583, Japan}
\newcommand{\stockholm}{The Oskar Klein Centre for Cosmoparticle Physics,
Department of Physics, Stockholm University, SE-106 91 Stockholm, Sweden}
\newcommand{\princeton}{Department of Physics: Joseph Henry Laboratories, 
Jadwin Hall, Princeton University, 
Princeton, New Jersey 08542, USA}
\newcommand{\VSI}{Van Swinderen Institute for Particle Physics and Gravity,\\ University of Groningen,
Nijenborgh 4, 9747 AG Groningen, The Netherlands}
\newcommand{\kavli}{Institute of Astronomy and Kavli Institute for Cosmology, Madingley Road, Cambridge, UK, CB3 0HA}
\newcommand{\damtp}{DAMTP, Centre for Mathematical Sciences, Wilberforce Road, Cambridge, UK, CB3 0WA}
\newcommand{\imperial}{Physics Department, Imperial College London, Prince Consort Road, London SW7 2AZ, UK}
\newcommand{\cita}{Canadian Institute of Theoretical Astrophysics, 60 St George St, Toronto, ON M5S 3H8, Canada.}
\newcommand{\ASU}{School of Earth and Space Exploration, Arizona State University, Tempe, AZ, 85287, USA}

\author{William R. Coulton}
\affiliation{\kavli}
\affiliation{\IPMU}
\author{Benjamin Beringue}
\affiliation{\damtp}
\author{P.\ Daniel Meerburg}
\affiliation{\VSI}

\date{\today}

\begin{abstract}
Anisotropies in the cosmic microwave background (CMB) are primarily generated by Thomson scattering of photons by free electrons. Around the time of recombination, the Thomson scattering probability quickly diminishes as the free electrons combine with protons to form neutral hydrogen. This production of neutral hydrogen enables a new type of scattering to occur: Rayleigh scattering of photons by hydrogen atoms. Unlike Thomson scattering, Rayleigh scattering is frequency dependent resulting in the generation of anisotropies with a different spectral dependence. Unfortunately the Rayleigh scattering efficiency rapidly decreases with the expansion of the neutral universe, with the result that only a small percentage of photons are scattered off of neutral hydrogen after recombination. Although the effect is very small, future CMB missions with higher sensitivity and improved frequency coverage are poised to measure the effect of Rayleigh scattering. The uncorrelated component of the Rayleigh anisotropies contains unique information on the primordial perturbations that could potentially be leveraged to expand our knowledge of the early universe. In this paper we explore whether measurements of Rayleigh scattering anisotropies can be used to constrain primordial non-Gaussianity and examine the hints of anomalies found by WMAP and \textit{Planck} satellites. We show that the additional Rayleigh information has the potential to improve primordial non-Gaussianity constraints over pure Thompson constraints by $30\%$, or more. Primordial bispectra that are not of the local type benefit the most from these additional scatterings, which we attribute to the different scale dependence of the Rayleigh anisotropies. Unfortunately this different scaling means that Rayleigh measurements can not be used to constrain anomalies or features on large scales. On the other hand, anomalies that may persist to smaller scales, such as the potential power asymmetry seen in WMAP and \textit{Planck}, could be improved by the addition of Rayleigh measurements.
\end{abstract}

\pacs{Valid PACS appear here}
\maketitle

\section{Introduction}
Measurements of the cosmic microwave background (CMB) anisotropies have been one of the best means of studying the primordial universe as the CMB anisotropies are linearly related to the primordial fluctuations. Through these measurements we have found that the fluctuations generated in the early universe can be accurately described as adiabatic, isotropic and Gaussian with a simple power-law power spectrum \citep[see e.g.][]{planck2016-l10,planck2016-l07,planck2016-l09}. Whilst these measurements have already been highly informative for our understanding of the early universe, there is still a broad range of models that describe their potential origin. 
To distinguish these models we require new primordial signatures. One such signature would be the measurement of primordial gravitational waves, and is the focus of much on-going work \citep{Ade2019,Abazajian2016,BICEP2018,Suzuki2018,Hanany2019}. Here we are motivated by searches for violations of three of the primordial fluctuations' properties: deviations from Gaussianity, isotropy and power-law behaviour. Any such violations would be highly constraining for models of the early universe.

Searches for these violations have been rigorously performed and to date there is no conclusive evidence for any violations. However, there are some interesting features in the CMB including: a lack of large angle correlations \citep{Spergel2003,Bennett2003,Copi2015}, a lack of variance \citep{Monteserin2008}, a hemispherical (or dipolar) power asymmetry \citep{Eriksen2004,Hansen2009}, a preference for power in odd multipoles \citep{Land2005},  weak hints of a feature in the CMB power spectrum \citep{planck2014-a24,Spergel2003} , an alignment of quadrupole and octopole \citep{Tegmark2003} and an anomalous cold spot \citep{Cruz2005,Vielva2004}. The significance of most of these anomalies is mild $2-3 \sigma$ and is complicated by the role of a posteriori choices  \citep{Bennett2011}, however adding new measurements would either clarify their origin as statistical fluctuations or point to new physics. Further, whilst searches for non-Gaussianity have found no significant deviations \citep{planck2016-l09}, a detection of primordial non-Gaussianity would be highly informative for our understanding about the physics in the early universe and our constraints are still much larger than theoretically interesting thresholds. Both of which motivate searching for methods to reach beyond current observational constraints \citep[see][for a summary of the current status]{Meerburg2019}.

It is challenging to improve our measurements of these effects. Hints of violations of isotropy and the power-law structure of initial fluctuations seem to occur at very large scales where cosmic variance means that there is limited room to improve our measurements with the CMB. Either the upcoming space based CMB mission LiteBird or the proposed satellite  Probe of Inflation and Cosmic Origins (PICO) would be able to provide only small improvements on existing large scale measurements and such measurements would reach the cosmic variance limit from the CMB \citep{Hanany2019,Suzuki2018}. At the same time, these scales are very difficult to measure with large scale structure surveys (which probe much smaller scales) \citep{Bhuvnesh2010,Dore2014}. To improve our constraints of non-Gaussianity we need to measure more modes. Upcoming CMB experiments will measure new small scale modes and this will lead to improved constraints \citep{Abazajian2016,Ade2019,Maresuke2019}. However it will be challenging to extract primary CMB modes down to very small scales due to Galactic and extra-galactic foregrounds and damping of primary fluctuation \citep{Silk1968}. Likewise, large galaxy and 21cm surveys can provide powerful constraints on some types of primordial non-Gaussianity  \citep{Dore2014,Alvarez2014,Dodelson2016,Munchmeyer2019,Castorina2020}.  Unfortunately, for many types of non-Gaussianity it is challenging to constrain these with large scale structure surveys due to signal contamination from the non-Gaussian evolution of gravity and structure formation, as well as observational limitations and computationally expensive estimators \citep{Karagiannis2018,Karagiannis2019,Dizgah2020}. 

Given the challenges of existing observables we have been motivated to explore alternative measures. In particular, in this work we will consider anisotropies sourced by Rayleigh scattering. Rayleigh scattering is the scattering of low energy photons by neutral atoms, whereas Thomson scattering, the dominant scattering process in the early universe, is the scattering of photons from charged particles. During recombination the universe transitioned from ionised to neutral resulting in an abundance of neutral hydrogen, which then scattered the CMB photons inducing further anisotropies. The impact of Rayleigh scattering on the CMB was first considered by \citet{Peebles1970} and then \citet{Yu2001} explored how Rayleigh scattering would impact measurements of the Thomson anisotropies for Wilkinson Microwave Anisotropy Probe (WMAP) and the \textit{Planck} experiments.  More recently \citet{Lewis2013} and \citet{Alipour2015} revisited the topic and explored how measurements of Rayleigh anisotropies could improve our constraints on the parameters of $\Lambda$CDM. Most recently, \citet{beringue2020} has expanded this effort by exploring the Rayleigh signal to probe extensions to $\Lambda$CDM with upcoming CMB experiments. These studies were primarily aimed at estimating the discovery potential of the signal and parameter constraints from the `late' universe (though we note that \citet{Lewis2013} also discussed how Rayleigh scattering can be used to disambiguate primordial gravitational B modes and lensing B modes). Here we build on these results but focus on the physics of the primordial universe to explore whether measurements of Rayleigh anisotropies can provide more information on potential primordial large scale anomalies or non-Gaussianity.

In Section \ref{sec:rayIntro} we review the physics of Rayleigh scattering and its prospects for cosmology. We discuss what we can learn about deviations from a power-law spectrum of primordial fluctuations and violations of isotropy using Rayleigh scattering in Section \ref{sec:featuresIsotropies}. The rest of the paper is focused on using Rayleigh scattering measurements to improve our constraints on primordial non-Gaussianity. We first review the origin of primordial non-Gaussianity and why it is interesting in Section \ref{sec:introprimordial non-Gaussianity} before discussing, in Section \ref{sec:resultsprimordial non-Gaussianity}, how constraints on primordial non-Gaussianity are altered by including measurements of Rayleigh scattering. We conclude in Section \ref{sec:conclusions}.

\section{Cosmology and Rayleigh scattering} \label{sec:rayIntro}
In this section we review the physics of Rayleigh scattering processes and its imprint on the CMB. For a thorough discussion of cosmological Rayleigh scattering we refer the reader to \citet{Lewis2013} and \citet{Alipour2015}. Rayleigh scattering refers to the scattering of long wavelength photons off neutral atoms. The internal dipole of the particle is excited by the incoming electromagnetic wave and radiates in return, producing an apparent scattering.

Recombination of free electrons and protons around $z \sim 1100$ produces neutral species off which CMB photons that have just decoupled from the plasma can scatter. Contrary to Thomson scattering, Rayleigh scattering is a frequency dependent effect with the cross section described by 
\begin{align} \label{eq:crossSecRay}
\sigma_R(\nu) =  & \sigma_T \left[ \left(\frac{\nu}{\nu_{\rm eff}}\right)^4+\frac{638}{243}\left(\frac{\nu}{\nu_{\rm eff}}\right)^6 \right. \nonumber \\ & \left. +\frac{1299667}{236196}\left(\frac{\nu}{\nu_{\rm eff}}\right)^8+.... \right]
\end{align}
where $\nu$ is the radiation frequency, $\sigma_T$ is the Thomson scattering cross section and $\nu_{\rm eff}\sim 3.1 \times 10^6$GHz is approximately the Lyman limit frequency. The frequency dependence of the cross section means that Rayleigh scattering anisotropies have a different spectral dependence to Thomson anisotropies and are increasingly important at high frequencies. The frequency dependence also means that Rayleigh scattering is only important over a smaller redshift range (with the effect decreasing as $a^{-7}$, where $a^{-3}$ comes from density dilution and $a^{-4}$ from the frequency dependence). Rayleigh scattering can be included in Boltzmann codes by modifying the Thomson scattering term to include the above frequency dependence. In this work we use a version of CAMB \citep{Lewis1999,Lewis2002} developed in \citet{Lewis2013}.

\citet{Alipour2015} found that the inclusion of measurements of Rayleigh anisotropies can help to improve constraints on parameters of $\Lambda$CDM, while a recent study \citet{beringue2020} showed that some extensions of $\Lambda$CDM could also benefit if Rayleigh scattering is included in cosmological inference. The Rayleigh anisotropies are a small correction to the Thomson anisotropies as can be seen from Eq.~ \eqref{eq:crossSecRay} and the size of  $\nu_{\mathrm{eff}}$ compared to CMB frequencies. For example, at 545 GHz the Rayleigh temperature anisotropies are $\sim2\%$ the size of the Thomson anisotropies. There is no frequency where the Rayleigh anisotropies are the dominant sky signal and the main limitation to the use of Rayleigh information is how well Rayleigh anisotropies can be separated from the brighter foregrounds and the primary CMB. Particularly problematic will be dust anisotropies which are orders of magnitude brighter than the Rayleigh anisotropies at high frequencies. In this work we defer addressing this complex issue and we will assume that we have a foreground cleaned map (with experimental properties that are expanded on below). 

\section{Constraining large scale features and violations of isotropy} \label{sec:featuresIsotropies}

\begin{figure}[t]
  \centering
  \includegraphics[width=0.48\textwidth]{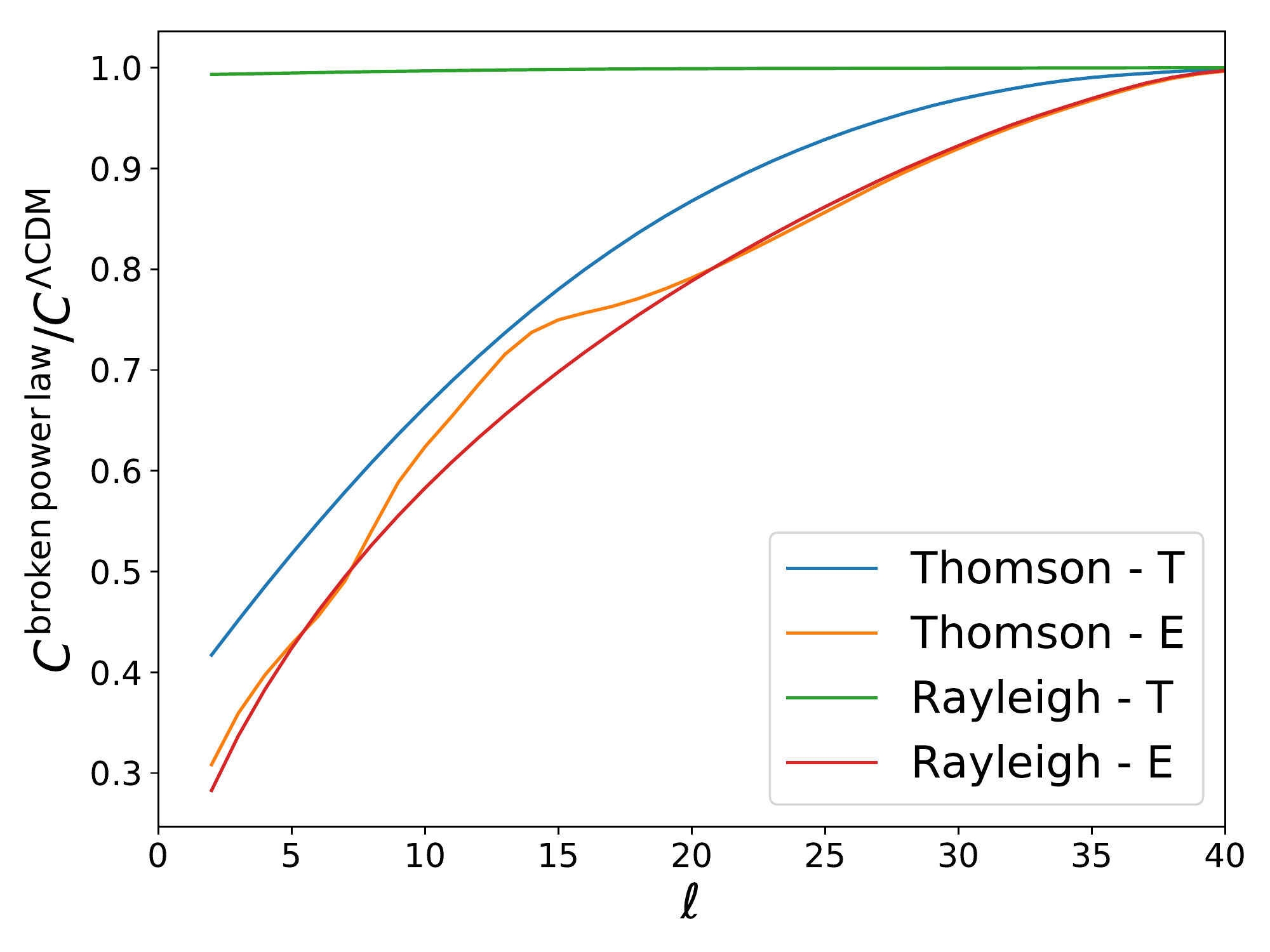} 
\caption{The ratio of the measured CMB Thomson and Rayleigh auto-spectra for a broken power-law primordial power spectrum,  given in Eq. \eqref{eq:brokenPowerlaw}, to the $\Lambda$CDM spectrum. We find that the Rayleigh temperature spectrum is unaffected by the large scale feature.}
\label{fig:brokenPS}
\end{figure}

The presence of large scale Rayleigh intensity and polarization anisotropies raises the questions of whether measurements of Rayleigh anisotropies can be used to study tentative anomalies on large scales. If these anisotropies probe similar scale modes, then there is the potential to double the statistical power. 

We wish to study the properties of the large scale primordial curvature perturbations, $\zeta(k)$, where $k$ is the comoving wavenumber. To investigate the sensitivity of Rayleigh anisotropies to these modes, we perform a simple test: we introduce a break in the primordial power spectrum so the primordial power spectrum, $\langle \zeta(k)\zeta(k') \rangle =P_{\zeta \zeta}(k) (2\pi)^3 \delta^{(3)}(k-k')$, is described as
\begin{align}\label{eq:brokenPowerlaw}
P_{\zeta \zeta}(k) = A_s &  \left( \mathcal{H}(k-k_{\rm{break}})  \left(\frac{k}{k_p}\right)^{n_s-4}   \right. \nonumber \\ 
&+\left. \mathcal{H} (k_{\rm{break}}-k) B \left(  \frac{k}{k_p}\right)^{n_b-4} \right),
\end{align}
where $ \mathcal{H}$ is the Heaviside function, $A_s$ is the amplitude of fluctuations, $n_s$ is the spectral tilt and $B$ is a constant such that the power spectrum is continuous. The power spectrum has a different spectral index, $n_b$ above the break scale, $k_{\rm{break}}$. In this test the break scale is chosen to be $k_{\rm{break}} = 0.005$ Mpc$^{-1}$ and we use a red spectral index of $n_b = 0.5$. In Fig.~\ref{fig:brokenPS} we plot the ratio of such a power spectrum to the $\Lambda$CDM power spectrum.  Interestingly we find that the large scale  Rayleigh temperature power spectrum is practically insensitive to the break in the power spectrum. This result is independent of the type of feature and holds for other modifications of the large scale power spectrum, e.g. for a localised features or sudden jumps. The effect on polarisation is larger, but the amplitude of the Rayleigh polarisation signal will be too small to be detected in the near future and hence we do not expect any statistical power to be obtained from these modes.

To understand the origin of this limited sensitivity it is insightful to investigate the various contributions to the Rayleigh spectrum. The evolution of the CMB anisotropies is governed by the Boltzmann equations which can be formally solved by the line-of-sight solution \citep{Seljak1996}. In the conformal-Newtonian gauge the line-of-sight solution can be written as
\begin{align} \label{eq:contributions}
\Theta_{\ell}(k,\eta_0,\nu) =  \int \limits_0^{\eta_0} \mathrm{d}\eta g(\eta,\nu) \left( \Psi (k,\eta)- \Theta_0(k,\eta) \right)j_{\ell}[k(\eta_0-\eta)] \nonumber \\
					+\int \limits_0^{\eta_0} \mathrm{d}\eta e^{-\tau(\eta,\nu)} \left(\dot{\Psi}(k,\eta)-\dot{\Phi}(k,\eta) \right) j_{\ell}[k(\eta_0-\eta)]   \nonumber \\
					-\int \limits_0^{\eta_0} \mathrm{d}\eta g(\eta,\nu) \frac{i v_b(k,\eta)}{k} \frac{\mathrm{d}}{\mathrm{d}\eta} j_{\ell}[k(\eta_0-\eta)]  \nonumber \\
					-\int \limits_0^{\eta_0} \mathrm{d}\eta\left(  \frac{1}{4} g(\eta,\nu) \Pi + \frac{3}{4k^2}\frac{\mathrm{d}}{\mathrm{d}^2\eta}\left(   g(\eta,\nu)\Pi \right) \right)j_{\ell}[k(\eta_0-\eta)], 
\end{align}
where $\Theta_{\ell}(k,\eta_0,\nu)$ is the fractional temperature perturbation, $g(\eta,\nu)$ is the visibility function, $\Psi (k,\eta)$ and $\Phi (k,\eta)$ are the metric potentials, $\tau$ is the optical depth, $\eta$ is conformal time with $\eta_0$ denoting the current conformal time, $v_b$ is the baryon velocity, $j_{\ell}[x]$ is the spherical Bessel function arising from the projection on the sphere, and $\Pi$ is the polarization tensor. The terms are grouped with the monopole sources in the first line, the integrated Sachs-Wolf (ISW) terms in the second, the Doppler sources in the third and the quadrupole sources in the final line. The quadrupole sources are small and so can be neglected for this discussion. Also note that there will be no late time ISW contribution for the Rayleigh spectrum as there is no `Rayleigh monopole' term.  

\begin{figure*}
\subfloat[Thomson components ]{
 \centering
  \includegraphics[width=.45\textwidth]{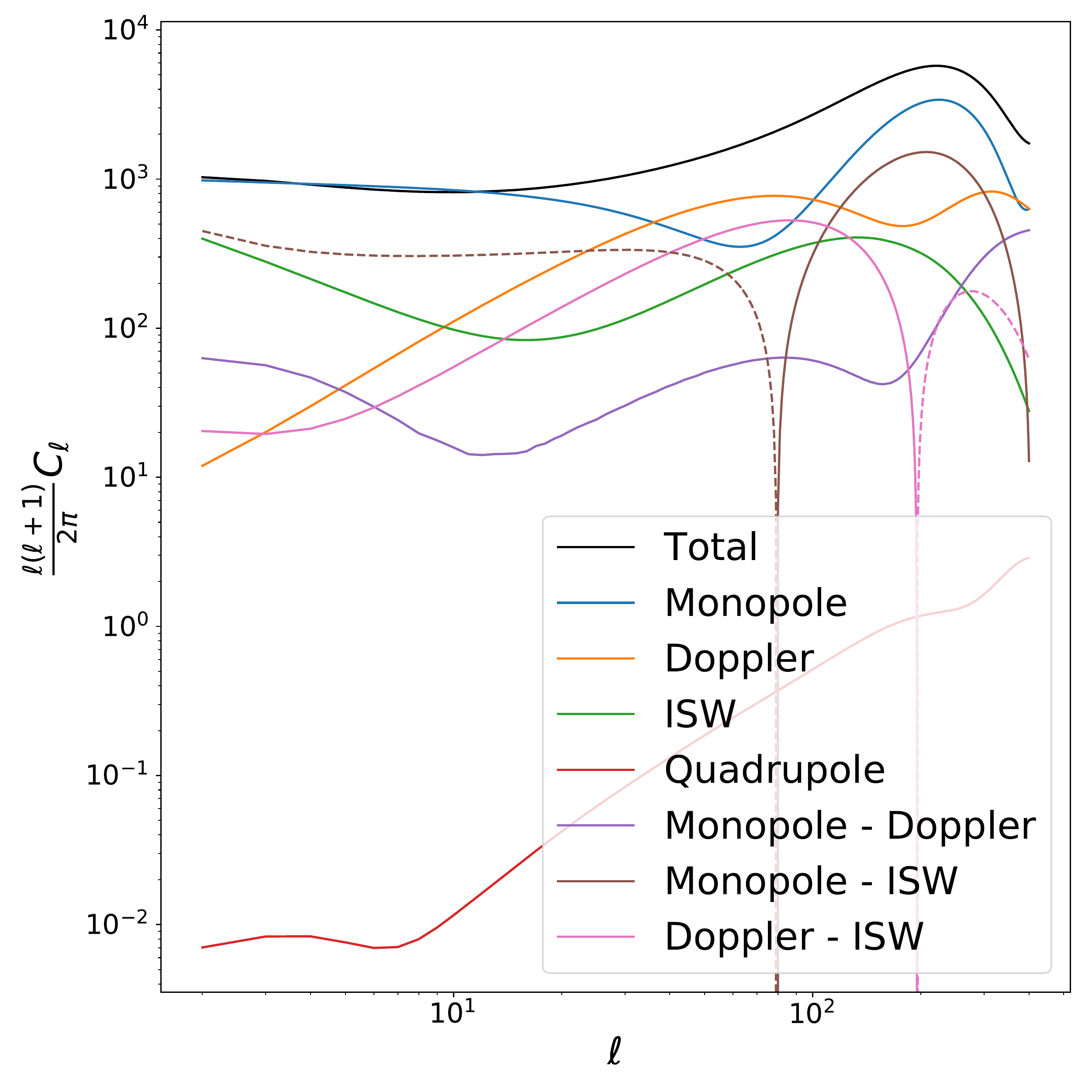}
  \label{fig:ThomsonComps}}
  \qquad
\subfloat[Rayleigh components ]{
  \centering
  \includegraphics[width=.45\textwidth]{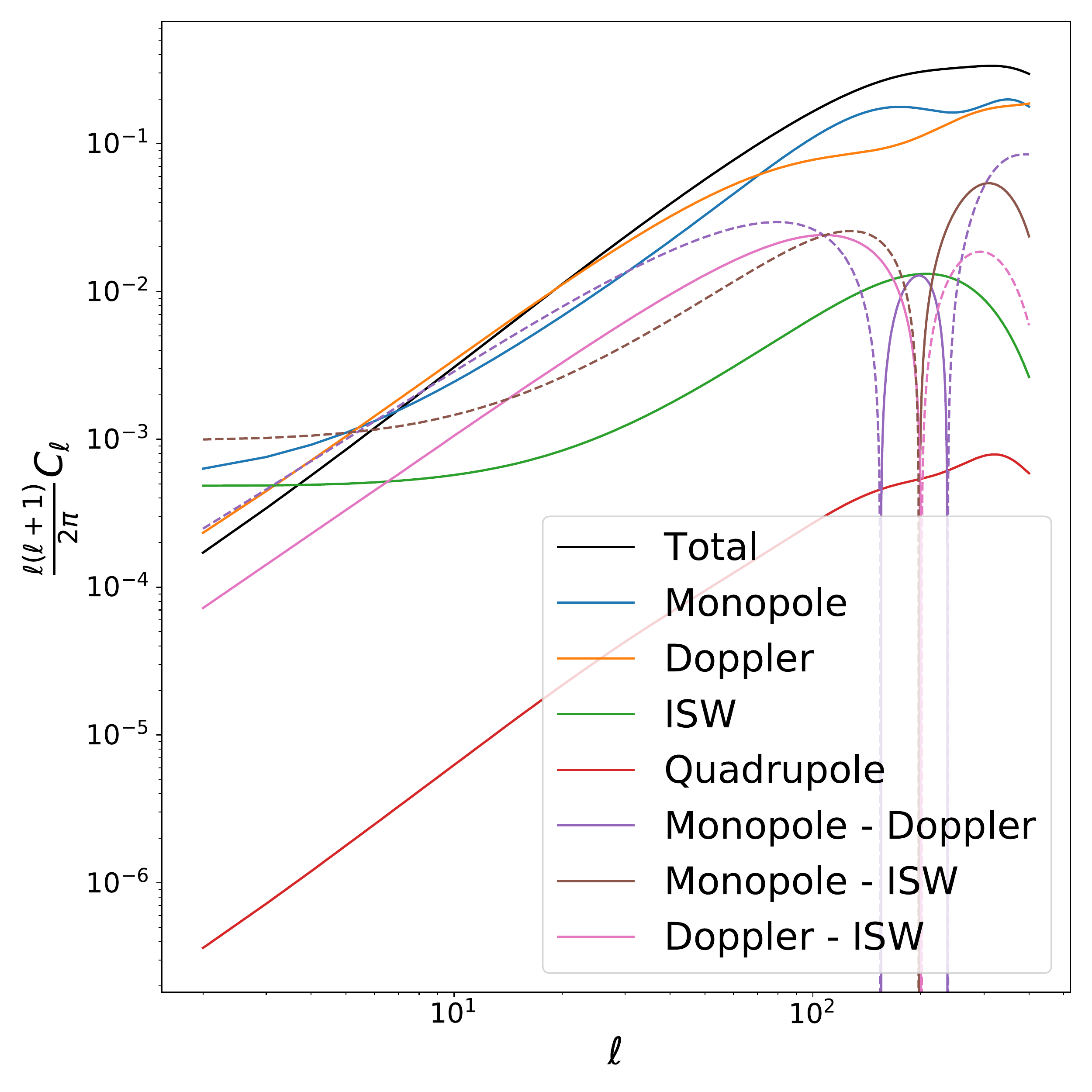}
  \label{fig:RayComps}}
\caption{The contributions to Thomson and Rayleigh temperature power spectrum from the different source terms including both the auto and cross terms. The cross terms include a factor of two as this is their contribution to the total power spectrum. Dashed lines denote negative values. Note that we do not show the contributions arising from cross correlations with the quadrupole as these are highly subdominant.}
\end{figure*}

Now consider the case of modes larger than the width of the last scattering surface. In this regime the Bessel function and monopole source terms are constant over the support of the visibility and can be pulled out of the integral. Utilising the condition that photons have scattered at some point in the history of the universe i.e.
\begin{equation} \label{eq:visib}
\int \limits_0^{\eta_0} \mathrm{d}\eta \, g(\eta,\nu) =1,
\end{equation}  
we obtain the Sachs-Wolfe result \citep{Sachs1967} that the large scale anisotropies are constant (if one neglects the Doppler sources due to their subdominant scaling as $k$). However we can now see that the visibility constraint, Eq.~\eqref{eq:visib}, is true at all frequencies and thus implies there are no anisotropies with Rayleigh frequency dependence from large scale modes. This means that the Rayleigh spectrum will have significantly smaller contributions from the monopole term and will only gain contributions when the modes vary on scales smaller than the width of the visibility function. In Fig.~\ref{fig:RayComps} we plot the individual contributions to the Rayleigh power spectrum. For comparison the equivalent contributions to the Thomson power spectrum are shown in Fig.~\ref{fig:ThomsonComps}.  We see that the monopole terms are significantly suppressed for Rayleigh scattering, especially when compared to the Thomson case. Physically Rayleigh scattering is a sub-horizon process and thus should not probe modes with scales larger than the horizon. 

As the Rayleigh temperature spectrum is insensitive to the large scale primordial power spectrum it is unsuitable to probe most of the potential large scale anomalies or features.  However, the Rayleigh spectrum could still be used to probe the dipole power asymmetry. As was seen in both WMAP \citep{Hansen2009} and \textit{Planck} \citep{planck2016-l07} the dipole power asymmetry does not seem to be limited to large scales (or temperature alone). Testing the Rayleigh anisotropies via methods such as the angular clustering of the power spectra on different scales \citep{planck2016-l07,planck2013-p09} would be especially interesting due to the different modes and scales probed.

We note that the Rayleigh $E$-mode spectrum is sensitive to large scale modes, as it is sourced by the large scale temperature quadrupole and, unlike the Thomson $E$-mode spectrum, only has contributions from recombination (there is no contribution from reionization). Thus it could be used to enhance the statistical power of isotropy tests and constraints on the large scale power spectrum. As an example we perform a simple test to examine the improvement in the constraint of the large scale spectral index, n$_b$, in our broken power spectrum model, Eq. \ref{eq:brokenPowerlaw}. We jointly constrain A$_s$, n$_s$ and n$_b$ finding that, in the CV limit, adding Rayleigh scattering tightens the constrain on n$_b$ by $15\%$ (note that this example is illustrative of the information content and a more thorough analysis should vary all the cosmological parameters, see e.g. \citep{Handley2019} ). As the Rayleigh $E$-mode signal is very small and significantly beyond any current or proposed experiment,  we do not further pursue a more detailed analysis of this direction in this work. 

Interestingly, this sensitivity to different scales has consequences for searches for non-Gaussianity, which is the focus of the remainder of the paper.

\section{Primordial non-Gaussianity}\label{sec:introprimordial non-Gaussianity}
Measurements of primordial non-Gaussianity are a powerful probe of the physics of the early universe. In this work we will focus on constraints on the primordial bispectrum, which is the Fourier equivalent of the three-point function. The bispectrum is the lowest order term in the Edgeworth expansion around a Gaussian PDF \citep{Babich2005} (i.e. the lowest order deviation from a Gaussian distribution) and for many primordial mechanisms this is the most sensitive statistic to deviations from Gaussianity. The primordial bispectrum, $B(k_1,k_2,k_3)$, is defined as
\begin{equation}
\langle \zeta(\mathbf{k}_1) \zeta(\mathbf{k}_2) \zeta(\mathbf{k}_3) \rangle = (2\pi)^3 \delta^{(3)}(\mathbf{k}_1+\mathbf{k}_2+\mathbf{k}_3)  \frac{3}{5} B(k_1,k_2,k_3).
\end{equation}
Note that the factor of $ \frac{3}{5} $ is included so that our definition is consistent with the convention used in the literature. 
The simplest inflationary model (single-field slow-roll, SFSR, inflation) generates Gaussian fluctuations with only non-observably small corrections coming from the weak coupling to gravity. As a result of this general prediction, a measurement of primordial non-Gaussianity would therefore exclude such a model as the origin of structure in the universe. In addition, besides a detectable level of non-Gaussianity, the bispectrum has a shape and the shape is highly informative about the dynamics that played a role in the early universe.
The importance of primordial non-Gaussianitys as an empirical discriminator of early universe models has led to a broad interest in primordial non-Gaussianitys (see \citet{Chen2010,Meerburg2019} and references therein, for detailed discussions) and several important results have been established. Firstly, consistency relations have shown that all inflationary models driven by a single-field slowly rolling down its potential have bispectra that are slow roll suppressed in the squeezed limit, when  $k_1\ll k_2,k_3$ \citep{Maldacena2003,Creminelli2004,Pajer2013}. Therefore a discovery of this type of non-Gaussianity would rule out almost all SFSR models. 

Secondly, several generic, and well motivated, mechanisms have been found that lead to measurable levels of non-Gaussianity.  For example, in early universe models were multiple (scalar) fields play a role in the production of the fluctuations, primordial non-Gaussianitys of the local type are produced. Local primordial non-Gaussianitys can be described by the following bispectrum
\begin{align}
 B^{\rm local}(k_1,k_2,k_3) =  & 2 f^{\rm local}_{\mathrm{NL}}A_s^2k_p^{8-2n_s} \left[ \frac{1}{k_1^{4-n_s}k_2^{4-n_s}}+ \right. \nonumber \\  &\left. \frac{1}{k_2^{4-n_s}k_3^{4-n_s}}+\frac{1}{k_1^{4-n_s}k_3^{4-n_s}} \right].
\end{align}
Generically models with strong non-linear dynamics during inflation, leading to propagation of fluctuations with $c_s \neq 1$, generate equilateral and orthogonal type non-Gaussianity \citep{Creminelli2006,Senatore2010}, with bispectra
\begin{align}
&B^{\rm{ equil}}(k_1,k_2,k_3) = 6 A_s^2 k_p^{8-2n_s} f^{\rm {equil}}_{\mathrm{NL}}\left(-\frac{1}{k_1^{4-n_s}k_2^{4-n_s}}  \right. \nonumber \\ &\left.  - \frac{1}{k_2^{4-n_s}k_3^{4-n_s}}-\frac{1}{k_1^{4-n_s}k_3^{4-n_s} }-\frac{2}{(k_1k_2k_3)^{2(4-n_s)/3} } \right. \nonumber \\ & \left.+\left[(\frac{1}{k_1^{(4-n_s)/3}k_2^{(4-n_s)/3}k_3^{2(4-n_s)/3}} +\mathrm{5 perm.}\right] \right),
\end{align}
and 
\begin{align}
&B^{\rm{ orth}}(k_1,k_2,k_3) = 6 A_s^2 k_p^{8-2n_s} f^{\rm {orth}}_{\mathrm{NL}}\left(-\frac{3}{k_1^{4-n_s}k_2^{4-n_s}} \right. \nonumber \\ &\left. - \frac{3}{k_2^{4-n_s}k_3^{4-n_s}} -\frac{3}{k_1^{4-n_s}k_3^{4-n_s} }-\frac{8}{(k_1k_2k_3)^{2(4-n_s)/3} } \right. \nonumber \\ & \left.+\left[ \frac{3}{k_1^{(4-n_s)/3}k_2^{(4-n_s)/3}k_3^{2(4-n_s)/3}} +\mathrm{5 perm.}\right] \right).
\end{align}
Whilst we focus on these three types of non-Gaussianity there are many other interesting types, e.g. those arising from non-bunch Davies initial conditions \citep{Meerburg2009} , features in the potential \citep{Chen2007,Adshead2012}, or from Gauge-inflation models \citep{Barnaby2011} to name a few, and  results presented in this paper should be relevant to searches for these types of non-Gaussianity as well. 

Whilst constraining primordial non-Gaussianity is generally computationally expensive (naively scaling as the number of modes/pixels cubed), there are several efficient estimators that have been developed e.g. \citet{Bucher2016,Komatsu2005,Fergusson2010}. Constraints on primordial non-Gaussianity are usually expressed as constraints on the amplitudes of primordial bispectra templates (such as the local, equilateral and orthogonal templates discussed above). For CMB measurements these template amplitudes can be estimated by measuring the CMB bispectrum, which is related to the primordial bispectra by transfer functions \citep{Komatsu2001} as follows
\begin{align}
& \langle a^{X_1}_{\ell_1,m_1} a^{X_2}_{\ell_2,m_2} a^{X_3}_{\ell_3,m_3}\rangle  = \mathcal{G}^{m_1,m_2,m_3}_{\ell_1,\ell_2,\ell_3} b^{X_1,X_2,X_3}_{\ell_1,\ell_2,\ell_3},
\end{align}
where
\begin{align}
b^{X_1,X_2,X_3}_{\ell_1,\ell_2,\ell_3}  = \int r^2 \mathrm{d}r \prod \limits_i \int  \frac{2}{\pi}\mathrm{d}k_i g^{X_i}_T(k_i)j_{\ell_i}(k_i r)  B(k_1,k_2,k_3),
 \end{align}
 and $ \mathcal{G}^{m_1,m_2,m_3}_{\ell_1,\ell_2,\ell_3} $ is the Gaunt integral, $r$ is the comoving distance and $g^{X}_T(k)$ is the transfer function where $X$ denotes either the Temperature ($T$) or curl-free polarization ($E$) modes. In this work we focus on constraints on the amplitudes of the three primordial templates discussed above.

To date the best constraints on primordial non-Gaussianitys have come from measurements of the CMB bispectrum, with the leading constraints coming from the \textit{Planck} satellite \citep{planck2016-l09}. The CMB has proven a powerful observable to search for non-Gaussianity because the linear relationship between primordial fluctuations and CMB anisotropies preserves the primordial statistics. However, there is diminishing power to improve these constraints from the primary CMB anisotropies as the number of new modes that can be measured is limited due to the damping of the primary fluctuations, the presence of small scale secondary anisotropies that obscure the primary anisotropies, and the measurement challenges of pushing to very small scales \citep{Abazajian2016,Ade2019}. 
 
 A very promising avenue is to constrain primordial non-Gaussianity through measurements of the large scale structure (LSS). In principle LSS measurements have the power to dramatically improve the constraints as they measure many more modes (3D galaxy positions compared to the 2D CMB anisotropies) \citep{Dore2014,Alvarez2014,Dodelson2016}. Furthermore, LSS tracers are biased in a unique way in the presence of local primordial non-Gaussianitys \citep{McDonald2009,Dalal2008}. Whilst direct measurements of this bias are hindered by large scale cosmic variance, in principle through the use of multitracer analysis  \citep{Seljak2009} constraints on local primordial non-Gaussianity can possibly be improved by almost an order of magnitude this decade \citep{Schmittfull2018,Munchmeyer2019,Darwish:2020prn}. 
 
 Besides local primordial non-Gaussianitys, LSS constraints will rely on the galaxy bispectrum or trispectrum. Unfortunately non-linear evolution generates non-Gaussianity that can easily obscure the primordial signal. This obscuration, combined with uncertainties in the details of small scale physics, such as baryonic processes, and observational challenges, such as redshift uncertainites and foreground removal, mean that it challenging to improve significantly beyond current CMB measurements \citep{Liu_2014a,Liu_2014b,Parsons_2012a,Parsons_2012b,Kalus_2018,Ross_2012}. For example, \citet{Karagiannis2019} find that future generation 21cm experiments, such as the recently proposed PUMA experiment \citep{Castorina2020}, will only improve by a factor of two over current CMB constraints, all while practically mapping out all linear modes up to redshift 6. 
 
The limited anticipated improvement over current CMB constraints, strongly motivates a search for alternative observables. While the Rayleigh signal is small, it does not suffer from some of the uncertainties and challenges present in the LSS. The smallness of the Rayleigh signal will make it hard to detect, but the guaranteed signal is definitely within reach of upcoming CMB experiments and the potentially large foregrounds (discussed at the end of Section \ref{sec:rayIntro}) can generally be distinguished because of unique frequency dependence \citep{planck2014-a11,planck2016-l04} and are relatively small when considered in comparison to those that, for example, hinder all cosmological 21cm observations \citep{Liu_2009,Bernardi2009}.

\section{Results: primordial non-Gaussianity constraints with Rayleigh scattering}\label{sec:resultsprimordial non-Gaussianity}

\begin{figure*}
  \centering
  \includegraphics[width=\textwidth]{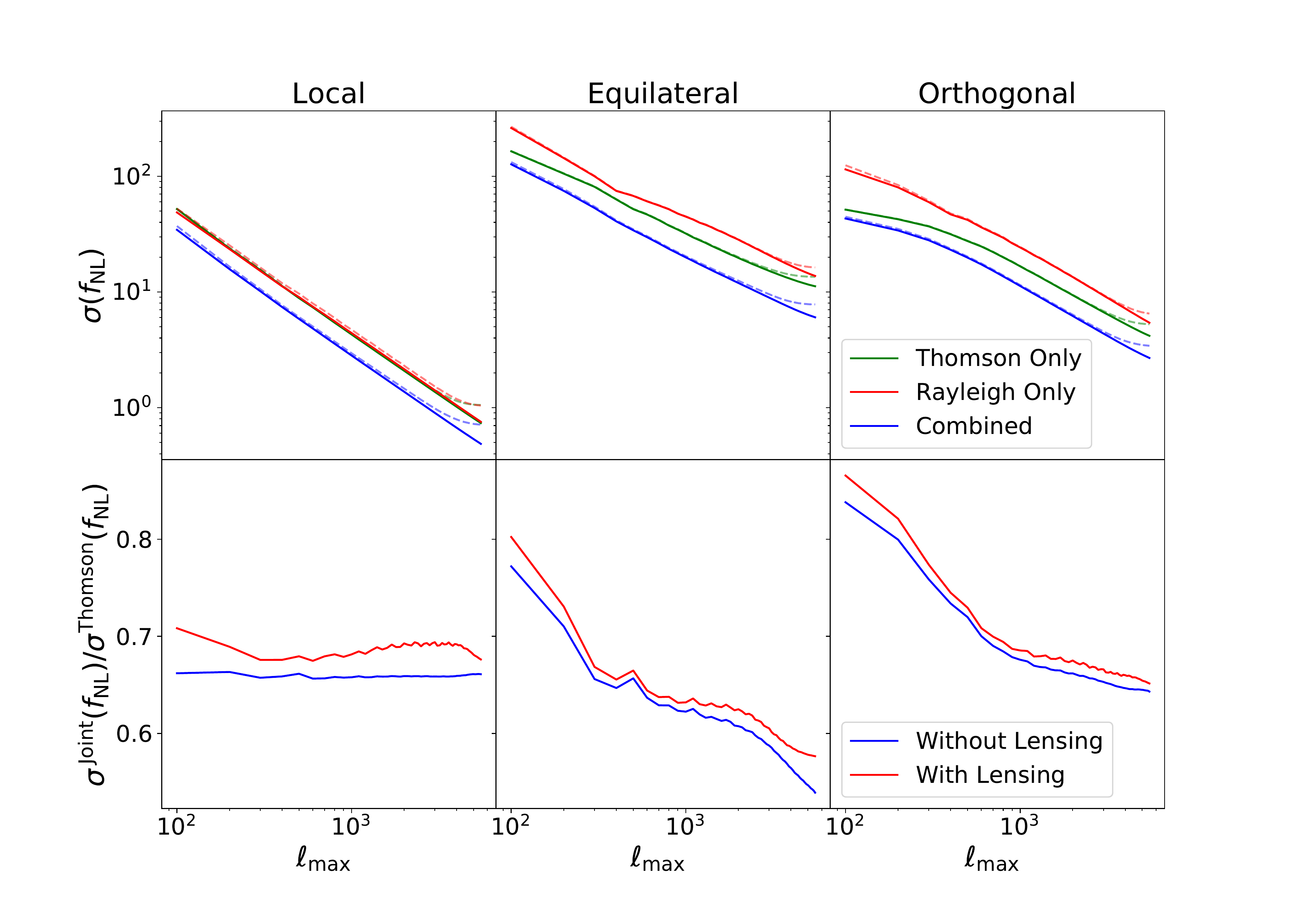} 
\caption{In the top panels we examine the cosmic variance limited constraining power of Thomson only (green), Rayleigh only (red) and combined (blue) measurements for three types of primordial non-Gaussianity as a function of the maximum scale used in the analysis. The dashed lines are the results when lensing effects are included in the power spectra. In the bottom panels we show the ratio of the joint Thomson and Rayleigh constraints compared to the Thomson only constraint. We find that adding Rayleigh measurements to Thomson measurements would improve the constraints on all types of non-Gaussianity by up to a factor of two.   }
\label{fig:cvLimit}
\end{figure*}

\begin{figure*}
  \centering
  \includegraphics[width=\textwidth]{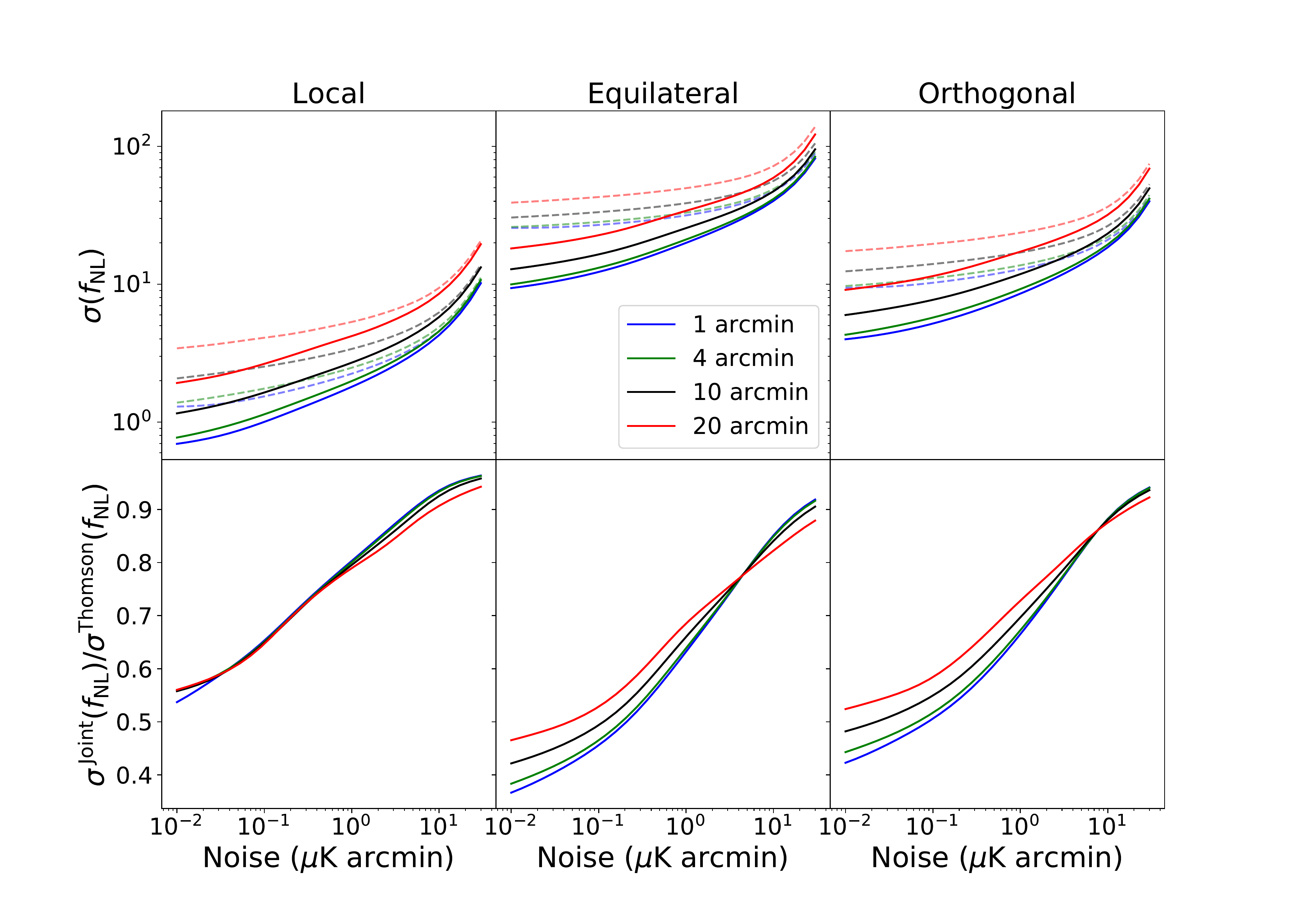} 
\caption{ In the top panels we explore how the constraining power of combined Thomson and Rayleigh (solid lines) and Thomson only measurements (dashed lines) varies as a function of instrument noise level and experimental beam for three types of primordial non-Gaussianity. The noise is treated as white and instrument properties are specified at  545 GHz, the observation frequency.  In the bottom panels we show the ratio of the joint Thomson and Rayleigh constraints compared to the Thomson only constraint for the equivalent experimental setup. Both sets of measurements include modes up to $\ell_{max} = 5500$. We see that experiments, which can isolate the Rayleigh signal with $\lesssim 5 \mu$K effective noise, obtain tighter constraints on primordial non-Gaussianity. We find that significant improvements can be obtained even for experiments with moderately large beams $\sim 4$ arcmin FWHM.  }
\label{fig:noiseAndBeam}
\end{figure*}

We use the Fisher forecast formalism to investigate whether Rayleigh scattering measurements can improve constraints on the amplitudes of local, equilateral and orthogonal types of non-Gaussianity. The predicted uncertainties are given by
\begin{align} \label{eq:fisherError}
&\frac{1}{\sigma^2(\hat{f}^{i}_{\mathrm{NL}})} = \frac{1}{6} \sum \limits_{\ell_i;X_i} \frac{(2\ell_1+1)(2\ell_2+1)(2\ell_3+1)}{4\pi} \begin{pmatrix}
   \ell_1 & \ell_3 & \ell_3 \\
   0 & 0 & 0 
  \end{pmatrix} ^2  \nonumber \\ &\times  b^{(i);X_1,X_2,X_3}_{\ell_1,\ell_2,\ell_3}{C^{-1}_{\ell_1}}^{X_1,X_4}{C^{-1}_{\ell_1}}^{X_2,X_5}{C^{-1}_{\ell_2}}^{X_3,X_6}b^{(i),X_4,X_5,X_6}_{\ell_1,\ell_2,\ell_3},
\end{align}
where ${C^{-1}_{\ell_1}}^{X,Y}$ is the observed power spectrum between map $X$ and $Y$. 

First, we consider the ideal case of a cosmic variance limited experiment; in this case the observed power spectra in  Eq.~\eqref{eq:fisherError} consist of the primary CMB alone. We consider constraints both including and excluding the effect of lensing on the power spectrum. We note that for configurations where this distinction matters, including the effect of lensing on the power-spectrum alone, will result in an over-estimation of the constraining power. \citet{Babich2004} and \citet{Coulton2019}  showed that in these regimes non-Gaussianity from lensing acts as extra noise on the bispectrum, increasing the noise beyond the level forecasted with Eq.~\eqref{eq:fisherError}. However, in practice this extra variance can be suppressed using delensing \citep{
Anderes2015, Larsen2016,Seljak2004,Smith2012,Green2017}, which if applied effectively, can mitigate this extra variance \citep{Coulton2019}. Note that lensing only impacts measurements on small scales ($\ell \gtrsim 2500$).

The constraining power of Rayleigh measurements in the CV case are shown in Fig.~\ref{fig:cvLimit}. Let us highlight the most interesting features.

First, we see that Rayleigh measurements on their own provide similar, but slightly worse, constraints on primordial non-Gaussianity compared to Thomson scattering measurements. Second, we find that combining Thomson and Rayleigh scattering improves constraining power, typically by $\sim 35\%$.  This improvement is consistent with an effective doubling of the number of modes. This is potentially surprising at first, given the strong correlations between the Rayleigh and Thomson anisotropies on small scales. However, this can be understood by considering a diagonalization of the observations into four uncorrelated parts (See e.g. the appendix of \cite{beringue2020}). Each of these uncorrelated components contains independent information on the primordial universe and inclusion of the Rayleigh signal should achieve an increase in constraining power consistent with the number of added modes. A key point to mention is that the removal of correlated component from, for example, the Rayleigh temperature mode, also removes the correlated contribution from the cosmic variance. Thus after diagonalization we are left with uncorrelated modes (with independent cosmological information) with variance limited by the cosmic variance of these modes alone.

\begin{figure*}[!t]
  \centering
  \includegraphics[width=\textwidth]{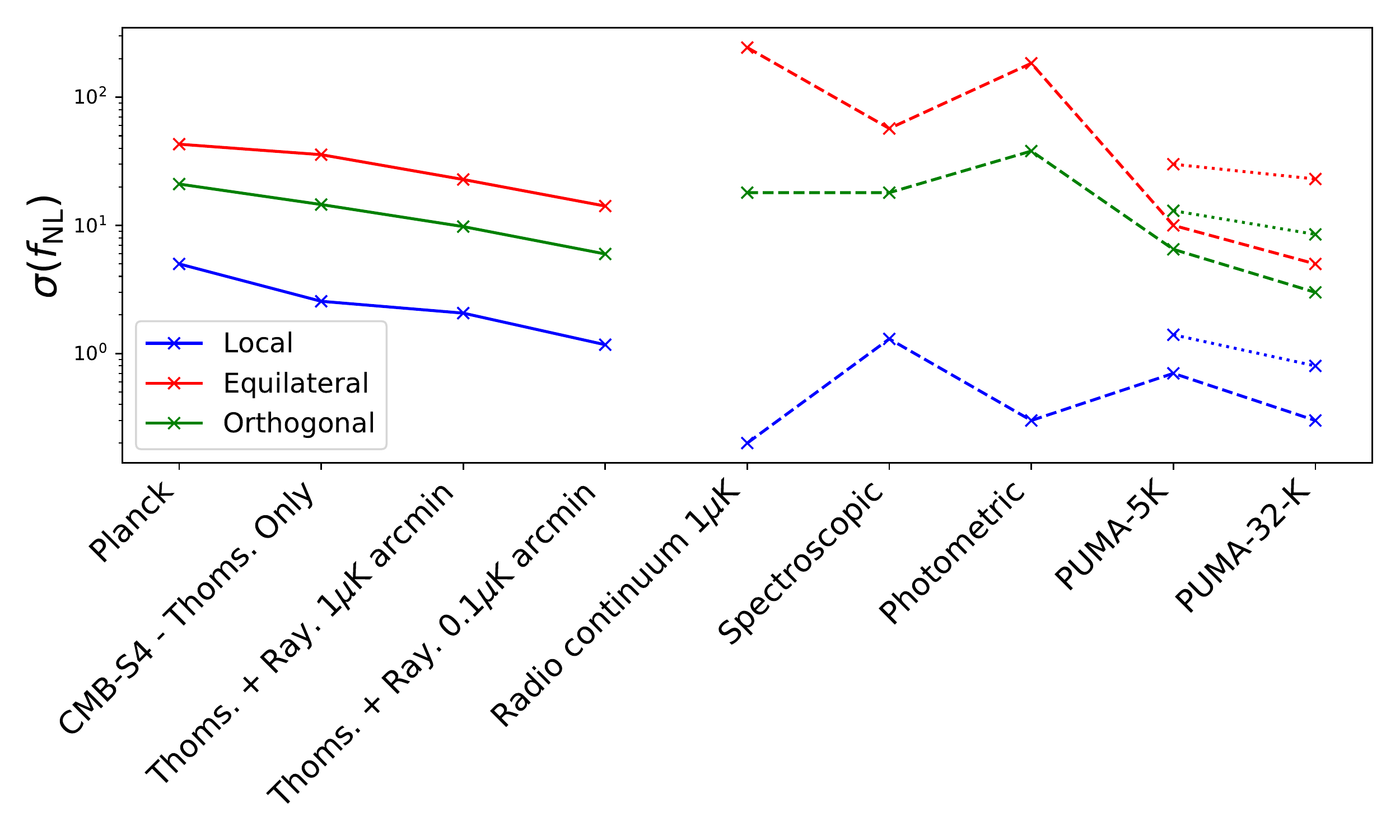} 
\caption{ A comparison of expected primordial non-Gaussianity constraints from CMB measurements (both with and without including Rayleigh measurements) and from a range of up-coming large scale structure surveys (LSS). We report the predictions for LSS surveys from \citet{Karagiannis2018,Karagiannis2019,Castorina2020}; the radio-continuum survey, the spectroscopic survey and the photometric survey are based on SKA-like, DESI-like and LSST-like surveys respectively. We see that the future CMB constraints, when Rayleigh measurements are included, provide competitive constraints to these upcoming LSS surveys. }
\label{fig:expComp}
\end{figure*}

Third, the relative improvement from adding Rayleigh scattering measurements shows some scale dependence and, for some templates, predicts an improvement that is {\it greater than just doubling the number of modes.} These results can be understood by recalling the discussion from Section \ref{sec:featuresIsotropies}. There we explained how the Rayleigh anisotropies are generally generated by different source terms and probe different scales compared to the Thomson anisotropies. It is this different coupling of scales that results in the template and scale dependence seen in Fig.~\ref{fig:cvLimit}.  For the equilateral shape we find a scale dependence that improves beyond the $1/\sqrt{2}$ improvement expected from doubling the number of modes. This improvement appears to be aiding a known problem. \citet{Bartolo2009} found that constraints on equilateral-type non-Gaussianity improve with $\ell_{\rm{max}}$ as  $1/\sigma(f^{\rm equil}_{\mathrm{NL}}) \propto \sqrt{\ell_{\rm max}}$, which is less than the expected improvement from mode counting arguments. This effect is due to the damping of the CMB on small scales\footnote{Although not the exponential damping, which actually cancels out. The signal-to-noise for $N$-point correlation function critically depends on their behaviour in the squeezed or collapsed limit. This will be explored in some depth in an upcoming publication (Kalaja et al.). See also \citet{Bordin:2019tyb} and \citet{Babich2004} for some further insights.} and to the finite width of the last scattering surface. Broadly, these effects result in a blurring of modes smaller that are smaller than either the damping scale or width of last scattering. This blurring effectively Gaussianizes the CMB signal.  Here we find that adding in Rayleigh scattering seems to restore some of the information loss. The constraints improve more rapidly with $\ell_{\rm max}$ closer to, but still less rapidly than, expectations from mode counting. This arises as the extra LSS surface of the Rayleigh signal undoes part of the bluring. Since the signal-to-noise for local non-Gaussianities already follows the mode-counting scaling for the Thomson signal, Rayleigh scattering does not have any effect on the ${\ell_{\rm max}}$ scaling. 

Since the Rayleigh signal appears to effectively double the number of measured modes, we will now explore a simplified experimental setup to investigate how observational effects impact these findings. Here we envisage a multifrequency experiment that has been used to clean out other sky signals, resulting in a map of the Rayleigh anisotropies at 545 GHz. We then explore the constraining power as a function of the instrument noise and beam in this 545 GHz map. The noise is assumed to be white. In this setup our observed power spectrum, in Eq.~\eqref{eq:fisherError}, consists of the unlensed CMB, the kinetic Sunyaev Zel'dovich effect (as this cannot be removed from the primary CMB Thomson anisotropies), instrument noise and beam effects. The kinetic Sunyaev Zel'dovich effect is assumed to have power $D_\ell= 3.0$ $\mu$K$^2$ at $\ell=3000$ consistent with recent measurements \citep{Reichardt2020}. We note that our results are similar if lensing is included in the power spectrum,  but note again that the dominant extra variance from the non-Gaussianity induced by lensing is not inlcuded here. In practice delensing would be used and this would result in constraints similar to as if the unlensed power spectrum is used \citep[see e.g.][for a more detailed discussion]{Coulton2019}.

In Fig.~\ref{fig:noiseAndBeam} we report our constraints on non-Gaussianity for this simplified experimental setup.  We find that that for noise levels below $\sim 10\; \mu$K arcmin Rayleigh measurements can improve over Thomson-only constraints and that for very low noise levels these constraints can be more than a factor of two better. Interestingly, for the equilateral and orthogonal constraints we find that we can gain significant improvements {\it even with low resolution experiments.} 

\section{Discussion and Conclusions}\label{sec:conclusions}

In this work we have explored whether measurements of Rayleigh anisotropies could be used to explore the initial conditions of the universe. Specifically we first investigated if including measurements of Rayleigh anisotropies can increase the statistical power of tests for deviations from isotropy or for large scale features in the primordial power spectrum. Our analysis found that, due to the suppresion of the non-Doppler source terms, the Rayleigh temperature power spectrum is insensitive to large angular scales, rendering it uninformative for the analysis of most (large scale) anomalies.  In principle the large scale Rayleigh $E$-mode spectrum is sensitive to anomalies on large angular scales, however this signal is likely too small to be detected in the near (or even distant) future.  We note that hints of the clustering of power spectra, which extends beyond large scales, as seen most recently in \citet{planck2016-l07} could be probed by Rayleigh anistropies and would present an interesting target.

In the remainder of this paper we focused on exploring how measurements of Rayleigh scattering can be used to improve constraints on primordial non-Gaussianity. We find that both in the noiseless case, and when assuming noise from a simplified experimental setup, Rayleigh measurements can provide significant improvements over projected constraints from the primary CMB signal alone. Including the Rayleigh signal can lead to improvements that are on par with futuristic 3D large scale structure surveys. For the noiseless case we find improvements of $\sim 35 \%$, consistent with doubling the number of observed modes. Interestingly even with experimental noise we find that we can tighten constraints by $\sim 40-50\%$ on the equilateral and orthogonal non-Gussianity, both of which are forecasted to be challenging to constrain with large scale structure surveys. 

It is important to point out that these improvements can be gained even by low resolution experiments. One of the main targets of future CMB experiments is to measure relic signatures from primordial gravitational waves \citep{BICEP2018,Crill2008,CLASS2014,Abazajian2016,Ade2019,Hanany2019,Suzuki2018}. To achieve this goal experiments are pushing to ever lower noise levels on large angular scales combined with a broad frequency coverage to remove potential contaminants. Such measurements, providing they extend to sufficiently high frequencies, could be well suited for measuring Rayleigh anisotropies and thus, as a byproduct of searches for signatures of primordial gravitational waves, we can potentially double our constraining power on primordial non-Gaussianity. 

We have not discussed any of the experimental challenges associated with isolating the Rayleigh signal from other (frequency dependent) sky signals. Disentangling the Rayleigh signal from Galactic dust and extragalactic emission from dusty star forming galaxies will be the biggest challenge. However removing Galactic dust is a similar concern for searches for signatures from primordial gravitational waves and so it is likely machinery to remove foregrounds contamination in the search for primordial gravitational waves can be equally applied to filter out the Rayleigh signal. The level to which we can clean the data without significant impact on the signal is still under active investigation.

Despite these potential challenges, detecting and utilizing the Rayleigh signal for cosmological inference is still very much worth the effort. This is convincingly shown in Fig.~\ref{fig:expComp} where we compare our projected Rayleigh constraints on primordial non-Gaussianity to predictions for proposed and upcoming photometric, spectroscopic, radio galaxy surveys as well as 21cm experiments \citep{Karagiannis2018,Karagiannis2019,Castorina2020}. While future large scale structre surveys will be able to significantly improve upon CMB constraints for local type non-Gaussianties, improving beyond current constraints on equilateral and orthogonal type non-Gaussianities is challenging and Rayleigh constraints are competitive with futuristic 21cm experiments for these shapes. Recent work by \citet{Watkinson2020} suggests that 21cm bispectrum measurements required to obtain these constraints could be even more challenging than was considered in \citet{Karagiannis2019}, because mixing between the foregrounds and the signal suggest there is no bispectrum equivalent of the power-spectrum's foreground wedge.  We stress that further work is required to fully assess the impact of this, particularly as \citet{Watkinson2020} considered studied the Epoch of Reionization, rather than the lower redshift regime $z \sim 0.3-6 $ relevant to  PUMA-like experiments. However, given these potential issues, the potential gains using the Rayleigh scattering signal to improve bounds on non-local non-Gaussian signals are evident.

\acknowledgements
The authors are very grateful for useful discussions with Anthony Challinor, Will Handley, Antony Lewis and  Joel Meyers. P.D.M. acknowledges support of the Netherlands organization for scientific research (NWO) VIDI grant (dossier 639.042.730). B.B. acknowledges support from the UK Science and Technology Facilities Council. W.R.C. acknowledges support from the UK Science and Technology Facilities Council (grant number ST/N000927/1). This work made use of the TIGER cluster at Princeton and used resources of the National Energy Research Scientific Computing Center (NERSC), a U.S. Department of Energy Office of Science User Facility operated under Contract No. DE-AC02-05CH11231.

\bibliographystyle{apsrev.bst}
\bibliography{LSSr,planck_bib}

\end{document}